\documentclass[10pt,letterpaper]{article}
\usepackage{opex3,cite}

\usepackage{amssymb,amsmath,epsfig}

\def\eq#1{Eq.~(\ref{#1})}
\def\fig#1{Fig.~\ref{#1}}

\begin{document}

\title{Transformation optics for cavity array metamaterials}

\author{James Q. Quach,$^1*$, Chun-Hsu Su,$^2$ and Andrew D. Greentree$^3$}

\address{$^1$ Centre for Quantum Computer Technology, School of Physics, The University of Melbourne, Victoria 3010, Australia\\
$^2$ Department of Infrastructure Engineering, The University of Melbourne, Victoria 3010, Australia\\
$^3$ Applied Physics, School of Applied Sciences, RMIT University, Victoria 3001, Australia}
\email{$^*$quach.james@gmail.com}

\begin{abstract*}
Cavity array metamaterials (CAMs), composed of optical microcavities in a lattice coupled via tight-binding interactions, represent a novel architecture for engineering metamaterials. Since the size of the CAMs' constituent elements are commensurate with the operating wavelength of the device, it cannot directly utilise classical transformation optics in the same way as traditional metamaterials. By directly transforming the internal geometry of the system, and locally tuning the permittivity between cavities,  we provide an alternative framework suitable for tight-binding implementations of metamaterials. We develop a CAM-based cloak as the case study.

\copyright 2013 Optical Society of America
\end{abstract*}




\section{Introduction}

Metamaterials, as an arrangement of artificial elements designed to achieve novel electromagnetic properties, represent a major advancement in material science. Typically constructed with conductive elements smaller than the operating wavelength, these structures have demonstrated novel electromagnetic behaviour, such as negative refraction~\cite{shelby01} and cloaking~\cite{schurig06}. Photonic crystals (PhCs) may also mimic metamaterial behaviour~\cite{luo03,urzhumov10,liang11}. Tight-binding models have recently been investigated as a novel metamaterial architecture~\cite{quach10,su10}. A realization of this is the cavity array metamaterial (CAM) -- an optical lattice composed of evanescently-coupled microcavities~\cite{quach10}. The quantum mechanical properties of the metamaterial are produced by coupling each cavity to single atoms that can be individually controlled. By tailoring the system's dispersion properties, it can show effects generally found in classical metamaterials, such as negative refraction, evanescent wave amplification, and anomalous waveguiding properties,  but with the distinct exception that it operates fundamentally at the quantum level. Tuning the atomic properties opens up the possibility of reconfigurable metamaterials with quantum properties. 

Pivotal in the development of metamaterials was the advent of transformation optics (TO) as a methodology for designing novel electromagnetic devices, paving the way to the realization of a cloaking device~\cite{schurig06}. Critically, because CAMs operate on a wavelength commensurate with the size of its constituent elements, they cannot directly utilise classical TO in the same way as traditional metamaterial systems. Here we develop a framework for TO using CAMs.


Traditional TO works by applying a covariant co-ordinate transformation to Maxwell's equations and then finding the required permittivity, $\epsilon$, and permeability, $\mu$, for light to follow geodesics in this transformed co-ordinate system which in the original co-ordinate system would be curved paths~\cite{leonhardt06,pendry06}. 
An attempt has been made to find a quantum equivalent that involves an invariant transformation of the Schr\"{o}dinger equation, predicting cloaking of matter waves~\cite{zhang08}. As a fundamental technique this work may find applications in systems such as ultracold rubidium atoms in optical lattices where the wavelength of the atom is many times greater than the lattice spacing so the system can be treated as an effective medium, although there are significant technical difficulties that need to be overcome first. 

\begin{figure}[tb]
\center
\includegraphics[width=.7\columnwidth]{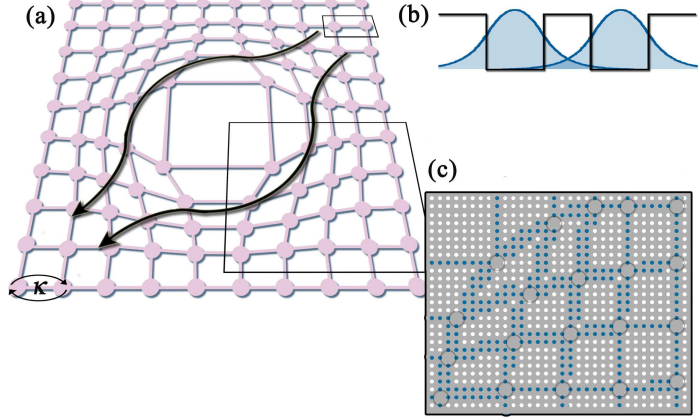}%
\caption{(Color online) (a) Schematic of a 2D array of coupled optical cavities that guides electromagnetic waves to form a cloaked region. (b) The inter-cavity coupling $\kappa$ can be achieved through evanescent field overlap. (c) An infiltrated silicon-based PhC. The refractive index of the liquid crystal infiltrated into air pores (blue) can be tuned with an applied electric field, providing a mechanism to dynamically control local intercavity couplings.}
\label{implementation_depictions}%
\end{figure}

PhCs as a metamaterial have the advantage that electromagnetic waves can propagate through them with relatively low loss. Classically, PhCs can effect TO by either varying the dielectric composition~\cite{urzhumov10} or geometrically transforming the PhC~\cite{liang11}. Our approach is in the spirit of the latter, in that we physically deform the medium. However our system differs from Ref.~\cite{liang11} because we are working with a quantum mechanical platform. Unlike Ref.~\cite{zhang08} we work with photons and not matter waves. In particular, we will look at the propagation of a single photon. 

In Sec.~\ref{sec:Cloaking} we discuss the guiding of light in CAMs to effect TO, using the cloaking of a circular region as a case study. We then  calculate in Sec.~\ref{sec:Permittivity detuning} the required permittivity distribution in a coupled optical array to implement the circular cloak.

\section{Cloaking}
\label{sec:Cloaking}

In CAMs, light propagation can be directly controlled by the lattice geometry and intersite hopping frequency. Therefore, unlike the continuous medium of classical metamaterials, co-ordinate transformations can be directly represented by changing the lattice configuration. The goal then is to keep the Hamiltonian invariant whilst the lattice configuration changes according to the required propagation response. Thus TO applied here to CAMs is a literal transformation of the medium space. 

To demonstrate TO in CAMs we consider an array of coupled optical cavities, the simplest of the CAMs, and show how it can be used as a cloak. This is schematically shown in \fig{implementation_depictions}(a). This choice of system, which has become increasingly accessible in experiments~\cite{notomi08}, suffices to demonstrate the effects of modifying geometry. The methods here are transferable to other more complex CAMs such as the Jaynes-Cummings-Hubbard (JCH) system~\cite{quach10}.

Consider a two-dimensional (2D) array of interacting identical cavities, described by the Hamiltonian ($\hbar=1$),
\begin{equation}
	\mathcal{H} = \omega\sum_i{a_i^\dagger a_i} - \kappa\sum_{\langle i, j \rangle}{ a_i^\dagger a_j},
\label{eq:hamiltonian}
\end{equation}
where $a_i^\dagger$ ($a_i$) is the photonic creation (annihilation) operator at site $i$, $\sum_{\langle i, j \rangle}$ is the sum over all nearest-neighbor cavities, $\omega$ is the cavity resonance frequency, and $\kappa$ is the inter-cavity coupling rate. We have used a second quantized description as it provides a convenient way to express the system, and it allows for a direct translation to more complex quantum mechanical systems, such as the JCH model. A classical electromagnetic description of the problem however, would be equally valid. Note that as the system is linear, and we are working explicitly with single photons, our results apply equally to many photons, where Hong-Ou-Mandel effects~\cite{hong87} are ignored.

\begin{figure}[tb]
\centering
\includegraphics[width=0.5\columnwidth]{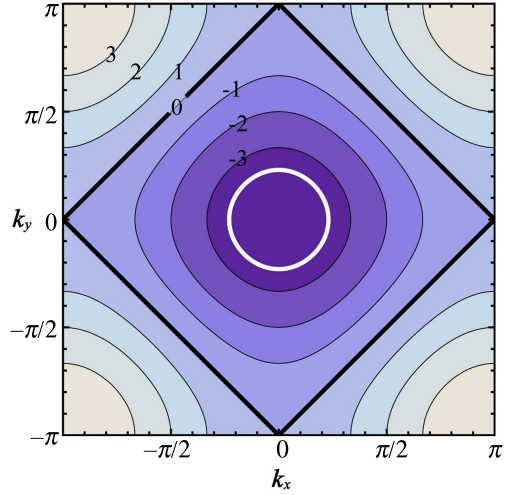}%
\caption{(Color online) Dispersion surface $E$ of a square coupled-cavity lattice system in the first Brillouin zone. The frequency of the contours are labeled with values of $(E - \omega)/\kappa$. The white contour indicates the operating frequency used to simulate the effective point source in \fig{simulation}(a). The black contour indicates the operating frequency used to simulate the beam source in \fig{simulation}(c).} 
\label{dispersion}%
\end{figure}

\begin{figure}[tb]
\center
\includegraphics[width=.7\columnwidth]{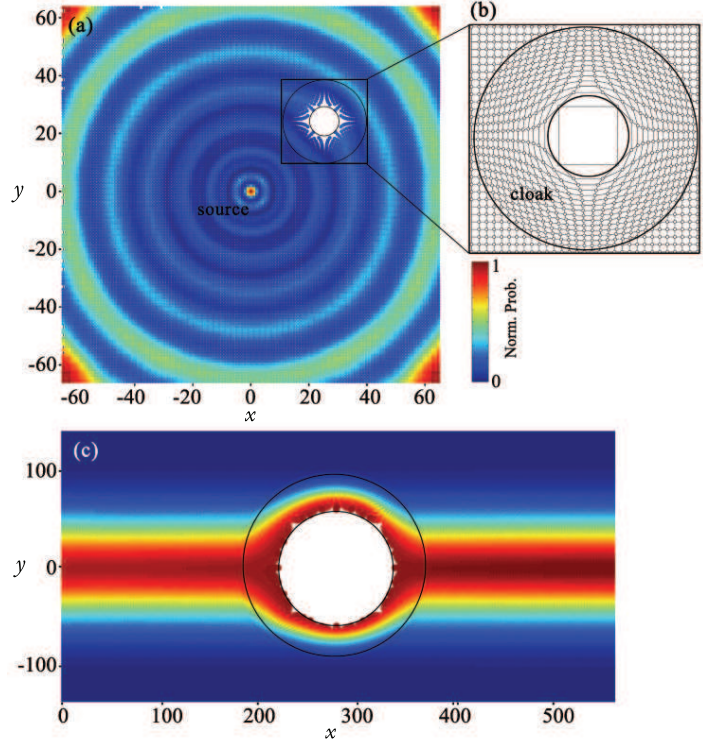}%
\caption{(Color online) (a) A quasi-point source forms a spherical wave. The spherical wave bends around the circular cloak region, rendering it effectively invisible. (b) The position of the cavity sites in the circular cloak are transformed according to \eq{eq:coordinate_transformation} with $a=5$ and $b=10$, making an annulus feature.  (c) Cloaking of a normally incident Gaussian pulse, superimposed in time to form a continuous beam. The circular cloak region is formed with $a=50, b=100$. The $x$ and $y$ axis indicate site co-ordinates.}
\label{simulation}%
\end{figure}

For a uniform square lattice, Bloch's theorem can be applied to find an analytical expression for the dispersion relation~\cite{quach09},
\begin{equation}
	E = \omega - 2\kappa[\cos(k_x d)+\cos(k_y d)]~,
\label{eq:energy_spectrum}
\end{equation}
where $\vec{k}=k_x \hat{x} + k_y \hat{y}$ is the wave vector associated with the \emph{crystal momentum} and $d$ is the distance between neighboring sites. The dispersion surface of the system is shown depicted in \fig{dispersion}. The behavior of the propagating modes is then governed by the group velocity,
\begin{equation}
			\vec{v}_g \equiv \nabla_{\vec{k}}E = 2\kappa d[\sin(k_x d) \hat{x} + \sin(k_y d) \hat{y}].
\label{eq:group_velocity}
\end{equation}
By spatially modifying the dispersion properties, or equivalently the shape of the isoenergy contours, of the array through the control of inter-cavity detuning or coupling, it has been shown that the light field can be guided to demonstrate unusual behaviors such as negative refraction~\cite{quach10} and superprism and collimating effects~\cite{su10}. In contrast, here we deform the array to affect the spatially varying dispersion properties so that the system acts as a circular cloak. We apply the standard coordinate transformation for such a cloak~\cite{pendry06}, i.e., in the new polar coordinates $(r',\phi')$,
\begin{equation}
	r' = \frac{b-a}{b}r + a,~~\phi' = \phi, 
\label{eq:coordinate_transformation}
\end{equation}
for the region $r \in (0,b)$, where $a$ is the radius of the hidden region and $b-a$ is the radial width of the cloak. The transformation directly relocates the relative positions of the cavities within this region, as shown in \fig{simulation}(b). 

To illustrate the cloaking property of the system, we solve the time evolution of the system, $\dot{\psi}(t) = -i\mathcal{H}\psi(t)$, in the one-excitation (single-photon) manifold, with two different initial conditions. We use isofrequency contours and group velocities given by Eq. (2) and (3) to identify operating frequencies. In \fig{simulation}(a), we initialize the excitation in a superposition of all the $k$-modes at an isofrequency contour near the band edge (white curve in \fig{dispersion}), as this is where it is most circular, to mimic a point-like source: we plot the probability distribution $|\psi(t)|^2$. In \fig{simulation}(c), the excitation is initialized as a Gaussian pulse in a single $k$-mode for a square contour (black bold curve in \fig{dispersion}) to minimize dispersion, and plot the superimposed distribution $\sum_{t}|\psi(t)|^2$. For both cases, the photon propagates around the hole and continues as if the hole was not there. Any detector outside the annulus region of deformation would not be able to distinguish between the original system and the cloaking system, thereby allowing anything inside the hole to be hidden from view. 

Unlike traditional transformation optics where a change in local $\epsilon$ and $\mu$ are required to effect the co-ordinate transformation, in CAMs the co-ordinate transformation is a direct transformation of the physical configuration of the system, and the inter-site properties are tuned to ensure the invariance of the Hamiltonian (Eq.~\ref{eq:hamiltonian}). In other words, the inter-cavity coupling for different inter-cavity spacing needs to remain the same for all sites $i$ and $j$ ($\kappa_{ij} = \kappa$, $\forall_{i,j}$). In the next section we show how this can be achieved by tuning the intercavity permittivity.

\section{Permittivity detuning}
\label{sec:Permittivity detuning}

Consider a 1D coupled cavity system with individual cavity width $w$, and permittivity profile described by step-functions such that $\epsilon(x) = \epsilon_a$ inside cavities and $\epsilon_b$ in-between cavities, as depicted in \fig{implementation_depictions}(b). Solving Maxwell's equation for a single cavity $\nabla^2 \textbf{E}_\Omega = -\epsilon(x)\omega^2 /c^2\textbf{E}_\Omega$ yields the fundamental cavity eigenmode with,
\small
\begin{equation}
|E_\Omega(x)| =\begin{cases}
	A\cos(\sqrt{\epsilon_a}\frac{\omega}{c} x), & |x| < w/2\\
	A\cos(\sqrt{\epsilon_a}\frac{\omega w}{2 c})e^{\sqrt{\epsilon_b}\frac{\omega}{c}(w/2 - |x|)}, & |x| > w/2\\
\end{cases}
\label{eq:cavity_eigenmode}
\end{equation}
\normalsize
where the normalization constant $A$ satisfies $\int{\epsilon_\Omega(x) |E_\Omega(x)|^2 dx} = 1$. In a tight-binding model, the nearest-neighbour coupling is given by~\cite{yariv99},
\begin{equation}
	\kappa = \frac{\omega}{2}(\epsilon_a-\epsilon_b)\int_{-w/2}^{w/2}{|E_\Omega(x)| |E_\Omega(x-d)|dx}.
\label{eq:hopping_frequency_1}
\end{equation}
To keep $\mathcal{H}$ invariant from lattice deformation, we solve \eq{eq:hopping_frequency_1} for a constant inter-cavity coupling while allowing $\epsilon_b$ to vary. 

Optical microcavities such as microdisks, pillars and toroids in general, and 2D PhC resonators in particular, represent a versatile platform for realizing this scheme. PhC cavities are created by introducing a local inhomogeneity in a periodic dielectric lattice. By placing such defects in close proximity, they become evanescently coupled to form a coupled-cavity array. Alternatively, this coupling can be achieved through a line-defect, which has demonstrated resonant transmission across over 100 cavities in a thin silicon PhC membrane~\cite{notomi08}. In both cases, the effective inter-site permittivity ($\epsilon_b$) is related to the refractive indices of the dielectric medium and hole infiltrations~\cite{busch99}, and also the holes' size, geometry and position~\cite{kuramochi06} in the crystal regions between the cavities. 

That the index of liquid crystal infiltrated into air pores of PhC can be altered substantially using an applied electric field~\cite{busch99} [\fig{implementation_depictions}(c)], represents a means to dynamically tune the local inter-cavity coupling. Mechanisms to change the index of the background material have also been developed including local heating~\cite{faraon09}, chemical etching~\cite{dalacu05}, electrical carrier injections~\cite{englund09}, near-field tip perturbation~\cite{gac09}, photosensitive material illumination~\cite{faraon08}, carbon dot deposition~\cite{seo08}, fiber-taper probing~\cite{shambat10}, ion-beam irradiation~\cite{tom09}, and nanoparticle intrusion~\cite{tomljenovic11}. Controlling the atomic-cavity detuning in coupled atom-optical cavities has also been shown to play a similar role to local index modification~\cite{quach10}. 

\begin{figure}[tb]
\center
\includegraphics[width=0.5\columnwidth]{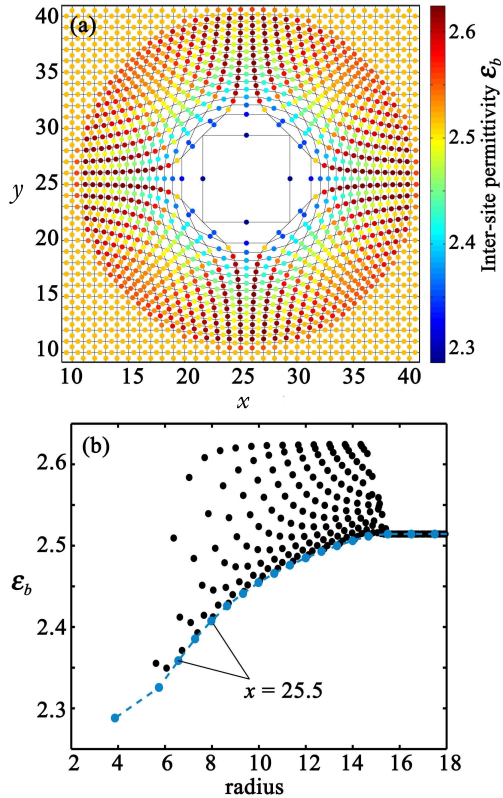}
\caption{(Color online) (a) Implementation of the cloaking device shown in \fig{simulation}, using a lattice distribution of inter-site permittivity $\epsilon_b$. $\epsilon_b$ values are denoted by the colored dots between the lattice sites. (b) Variation of the inter-site permittivity as a function of the distance from the cloaking center, in a quadrant (other quadrants are symmetrically equivalent). Near the centre, the intersite permittivity is lower than the baseline permittivity as the spacing between sites are stretched; whereas near the outer edge of the cloaking annulus the permittivity is higher as the sites are squashed closer together. The blue dots indicate the permittivity distribution along co-ordinate $x = 25.5$.}
\label{permittivity_and_detuning_profiles_1}%
\end{figure}

As a specific case study, we consider an infiltrated silicon-based PhC [\fig{implementation_depictions}(d)] implementation operating at $\lambda = 1.5~\mu$m with typical setup characteristics of $\epsilon_a = 11.7$, $\epsilon_b \approx 2.3$ (prior to transformation), $\kappa = 10^{14}$~rad/s, and $w = \lambda/2$. To achieve the above cloaking prescription with uniform $\omega$ and $\epsilon_a$, we require the distribution of the inter-site permittivity to take up values shown in \fig{permittivity_and_detuning_profiles_1} for the example in \fig{simulation}. The permittivity in-between neighboring sites is assumed to be constant. Each plot point in \fig{permittivity_and_detuning_profiles_1}b) represents an inter-site position, and is characterized by a radial distance and a permittivity value. Near the center of the cloak the deformation stretches the physical distance between nearest neighbors. To maintain the same hopping frequency, the intersite permittivity is decreased. Near the outer radius of the cloaking annulus, the sites are squashed closer together, and correspondingly the intersite permittivity here is greater than the baseline permittivity outside the cloaking region.

\section{Conclusion and outlook}
CAMs offers the potential of metamaterials with quantum properties. Here we have shown how transformation optics is implemented in the context of CAMs: a spatial transformation of the physical sites and careful tuning of the intersite coupling. As a case study, we showed how a single photon device can cloak a circular region. This work provides the framework for future investigation into quantum mechanical behaviour such as superposition and entanglement, in this system. These types of quantum mechancal behaviour could, for example, be achieved by coherently manipulating the states of atoms or atom-like systems coupled to the cavity sites. 

\section*{Acknowledgments}
We acknowledge useful discussions with Andrew. M. Martin, Andrew C. Hayward, and Lloyd C.L. Hollenberg. J.Q.Q. would like to thank Susan M. Heywood for support and general discussion. This work was supported by the Australian Research Council (ARC) under the Centre of Excellence scheme.  A.D.G. acknowledges the financial support of the ARC under Project DP0880466.


\begin{thebibliography}{99}

\bibitem{shelby01} R. Shelby,  D. R. Smith,  and S. Schultz,  ``Experimental verification of a negative index of refraction,'' Science {\bf 292,} 77--79 (2001).

\bibitem{schurig06} D. Schurig, J. J. Mock, B.J. Justice, S. A. Cummer, J. B. Pendry, A. F. Starr, and D. R. Smith, ``Metamaterial electromagnetic cloak at microwave frequencies,'' Science {\bf 314,} 977--980 (2006).

\bibitem{luo03} C. Luo, S. G. Johnson, J. D. Joannopoulos, and J. B. Pendry, ``Subwavelength imaging in photonic crystals,'' Phys. Rev. B {\bf 68,} 045115 (2003).

\bibitem{urzhumov10} Y. A. Urzhumov, and D. R. Smith, ``Transformation optics with photonic band gap media,'' Phys. Rev. Lett. {\bf 105}, 163901--163905 (2010).

\bibitem{liang11} Z. Liang, and J. Li, ``Scaling two-dimensional photonic crystals for transformation optics,'' Opt. Express {\bf 19} 16821--16829 (2011).

\bibitem{quach10} J. Q. Quach, C.-H. Su, A. M. Martin, A. D. Greentree, and L. C. L. Hollenberg, ``Reconfigurable quantum metamaterials'', Opt. Express {\bf 19}, 11018--11033 (2011).

\bibitem{su10} C.-H. Su, `` Novel quantum technology based on atom-cavity physics'', Ph.D. thesis, The University of Melbourne, Victoria (2010).

\bibitem{leonhardt06} U. Leonhardt, ``Optical conformal mapping'',Science {\bf 312,} 1777--1780 (2006).

\bibitem{pendry06} P. B. Pendry, D. Schurig, and D. R. Smith, ``Controlling electromagnetic fields'', Science {\bf 312,} 1780--1782 (2006).

\bibitem{zhang08} S. Zhang, D. A. Genov, C. Sun, and X. Zhang, ``Cloaking of matter waves'', \prl {\bf 100,} 123002 (2008).

\bibitem{notomi08} M. Notomi, E. Kuramochi, and T. Tanabe. ``Large-scale arrays of ultrahigh-Q coupled nanocavities'', Nat. Phot. {\bf 2,} 741--747, (2008).

\bibitem{hong87}  C. K. Hong, Z. Y. Ou, and L. Mandel.  ``Measurement of subpicosecond time intervals between two photons by interference'', \prl{\bf 59,} 2044--2046, (1987).

\bibitem{quach09} J. Quach, M. I. Makin, C.-H. Su, A. D. Greentree, and L. C. L. Hollenberg, ``Band structure, phase transitions and semiconductor analogs in one-dimensional solid light systems,'' \pra, {\bf 80,} 063838 (2009).

\bibitem{yariv99} A. Yariv, Y. Xu, R. K. Lee, and A. Scherer, ``Coupled-resonator optical waveguide: a proposal and analysis,'' Opt. Lett. {\bf 24,} 771-713 (1999).

\bibitem{busch99} K. Busch and S. John, ``Liquid-crystal photonic-band-gap materials: the tunable electromagnetic vacuum'' \prl {\bf 83,} 967 (1999). 

\bibitem{kuramochi06} E. Kuramochi, M. Notomi, S. Mitsugi, A. Shinya, T. Tanabe, and T. Watanabe, ``Ultrahigh-Q photonic crystal nanocavities realized by the local width modulation of a line defect,'' \apl {\bf 88}, 041112 (2006).

\bibitem{faraon09} A. Faraon and J. Vu{\v{c}}kovi{\'c}, ``Local temperature control of photonic crystal devices via micron-scale electrical heaters,'' \apl {\bf 95,} 043102 (2009).

\bibitem{dalacu05} D. Dalacu, S. Fr{\'e}d{\'e}rick, P. J. Poole, G. C. Aers, and R. L. Williams, ``Postfabrication fine-tuning of photonic crystal microcavities in InAs/InP quantum dot membranes,'' \apl {\bf 87,} 151107 (2005).

\bibitem{englund09} D. Englund, B. Ellis, E. Edwards, T. Sarmiento, J. S. Harris, D. A. B. Miller, and J. Vuckovic, ``Electrically controlled modulation in a photonic crystal nanocavity,'' Opt. Express {\bf 17,} 15409--15419  (2009).

\bibitem{gac09} G. Le Gac, A. Rahmani, C. Seassal, E. Picard , E. Hadji, and S. Callard, ``Tuning of an active photonic crystal cavity by an hybrid silica/silicon near-field probe,'' Opt. Express {\bf 17,} 21672--21679  (2009).

\bibitem{faraon08} A. Faraon, D. Englund, D. Bulla, B. Luther-Davies, B. J. Eggleton, N. Stoltz, P. Petroff, and J. Vuckovic, ``Local tuning of photonic crystal cavities using chalcogenide glasses,'' \apl {\bf 92,} 043123 (2008).

\bibitem{seo08} M.-K. Seo,  H.-G. Park, J.-K. Yang, J.-Y. Kim, S.-H. Kim, and Y.-H. Lee, ``Controlled sub-nanometer tuning of photonic crystal resonator by carbonaceous nano-dots,'' Opt. Express {\bf 16,} 9829--9837  (2008).

\bibitem{shambat10} G. Shambat, K. Rivoire, J. Lu, F. Hatami, and J. Vu{\v{c}}kovi{\'c}, ``Tunable-wavelength second harmonic generation from GaP photonic crystal cavities coupled to fiber tapers,'' Opt. Express {\bf 18,} 12176--12184  (2010).

\bibitem{tom09} S. Tomljenovic-Hanic, A. D. Greentree, C. M. de Sterke, and S. Prawer, ``Flexible design of ultrahigh-Q microcavities in diamond-based photonic crystal slabslexible design of ultrahigh-Q microcavities in diamond-based photonic crystal slabs,'' Opt. Express {\bf 18,} 6465--6475  (2009).

\bibitem{tomljenovic11} S. Tomljenovic-Hanic, A. D. Greentree, B. C. Gibson, T. J. Karle, and S. Prawer, ``Nanodiamond induced high-Q resonances in defect-free photonic crystal slabs,'' Opt. Express \textbf{19}, 22219--22226  (2011).


\end{thebibliography}
\end{document}